# The Stefan-Boltzmann constant re-visited for photons thermally generated within matter

G.B. Smith*, A.R. Gentle, M.D. Arnold, School of Mathematical and Physical Sciences, University of Technology Sydney; Broadway, NSW, Australia

*Corresponding author. Email: geoff.smith@uts.edu.au

**Abstract:** The Stefan-Boltzmann constant arose from photon densities inside a cavity, but inside matter photon mode densities are material specific. Photon speeds are governed by the mode they occupy, so mode densities can be expressed in terms of speed. Cavity intensities at temperature $T_K$ combined $[8\pi k^4/(c^3 h^3)]T_K^4$ with $(\pi^4/15)$. A material dependent number from summation of internal photon spectral energy densities replaces $\pi^4/15$. Spectral densities are presented for water, germanium and silver. Output intensity combines revised hemispherical emittance $\varepsilon_{Q,H}$ based on these densities, with universal factor $[8\pi k^4/(c^3 h^3)]T_K^4$. Emitted radiance after interface internal reflectance of directionally invariant internal radiance elements defines $\varepsilon_{Q,H}$. Predicted internal densities are verifiable using measured external spectral intensities, provided refraction upon exit is accounted for in emissivity, which the Kirchhoff rule neglects. "Virtual-bound state" photon resonances are predicted in dielectrics and observed.

**One sentence summary**: Spectral energy densities of photons within matter arise from their thermal occupation of material defined modes, so internal and refracted exit mode spectral intensities diverge from blackbody intensities.

**Main Text:**

Thermodynamics introduced us to photons through the work of Planck(*1-3*) on the internal energy density $U(T_K)$ and entropy flux density $S(T_K)$ in the interior of a closed cavity filled with radiation emerging from its walls at temperature $T_K$. He showed that electromagnetic energy consists of fluxes of photonic particles with energy (hf) with f the frequency of the stationary wave mode along which each particle moves. Planck found that $U(T_K)$ inside a cavity obeyed the relatively simple physical rule $U(T_K)= \sigma T_K^4$ Jm$^{-3}$. Cavity $\sigma$ consists of two numerical factors $[8\pi k^4/(c^3 h^3)]$ and $\pi^4/15$. This rule has taken on the status of a universal law meaning $\sigma$ is assumed to be a universal constant. This study shows that the number $\gamma = [8\pi k^4/(c^3 h^3)] = 8.7306 \times 10^{-9}$ Wm$^{-2}$K$^{-4}$ is the universal pre-factor for all internal and external densities, plus all total and directional external intensities, where it combines with $T_K^4$. The pre-factor $(\pi^4/15)\gamma$ occurs only for photons within a closed cavity where $(\pi^4/15)$ is also the hemispherical emittance $\varepsilon_H$ of a small aperture in a cavity wall. Inside matter $\pi^4/15$ is replaced by a number which is material and $T_K$ dependent. Example energy density spectra whose integration yields this new number will be presented for three materials. The material dependence of this replacement number, like Planck's derivation of cavity $\sigma$, arises from an integral over those ground-state quantum mode frequencies(*4*) which have been occupied according to Bose-Einstein weighting $1/[\exp(hf/kT_K)-1]$. The change to $U(T_K)$ is then derived from the link between photon mode density and speed $c^*(f)$ per mode. Photon density then varies between materials and at each frequency. The second section of this report defines a generalized quantum hemispherical emittance $\varepsilon_{Q,H}$ for all interiors based on matter dependent internal densities and interface internal directional reflectance, which vanishes at a cavity aperture.

Internal photons are excited thermally onto standing quantum modes within matter. These modes allow some photon transfer to modes created at the interface in the external continuum at energies matching those of impacting internal modes. These are created by the mean photon reaction potential at the exit interface acting on the internal incident mode. Internal photon modes thus share some characteristics with condensed matter ground state modes except that some of their photons can enter external continuum modes which have reduced amplitudes. Inside absorbers we will use the agreement between mean photon-based transport of power and Poynting vector power flows based on Maxwell's wave solutions. This requires that mean photon loss rates within an absorber duplicate the mean rate of power decay with time or distance travelled by an entering Maxwell wave. That equivalence allows mean transport properties of photons produced inside matter to be modelled using optical wave indices [n(f)+ik(f)] at frequency f. Internal photon speed c*(f) = c/n(f) is a modal characteristic at each frequency. Other photon properties within absorbers are discontinuous as annihilation and creation occurs in random collision events that involve a non-photonic excitation such as a phonon or excited electron. Classical amplitude and power decay is different being continuous in time and distance travelled. For photons time elapsed and distance travelled since each one's creation are thus variable as photon lifetimes τ(f) and distances travelled d*(f) are not fixed. Random photon creation and annihilation means that photon internal number density N($T_K$,f) fluctuates. The Bose-Einstein occupation factor thus describes the time-averaged occupation of each mode, while lifetime τ(f) and distance travelled d*(f) from creation per photon are variables with mean values.

Classical amplitude decay means power decay P(d*) with distance travelled d* into a lossy medium are described with an attenuation co-efficient α(f) which is frequency dependent so P(d*)=P(0)[exp(-α(f)d*)]. α(f) is described by the well-known optical relation α(f) = 4πk(f)f/c*(f) = 4πk(f)/λ*(f) with λ*(f) = λ/n(f). The mean distance a decaying classical internal power flow travels prior to full extinction <d(f)*>$_{Cl}$ can be found by integration of d* weighted by the power surviving at d* as in equation (1). σ(f) is optical conductivity, δ(f) skin depth and $\varepsilon_2$(f) =2n(f)k(f). Skin depth is thus the mean optical path of decaying classical wave power from its entry points into an absorber.

$$\langle d^*(f)\rangle_{Cl} = \int_0^\infty d(d^*)d^*(\exp(-\alpha(f)d^*)) = \frac{1}{\alpha(f)} = \frac{c}{\varepsilon_2 2\pi f} = \frac{c}{\sigma(f)} = \frac{\delta(f)}{n(f)} \qquad (1)$$

Classical power attenuation thus becomes P(d*)= P(0)[exp(-d*/(<d*(f)>$_{Cl}$)]. This generic relation was originally a survival equation(4) for a beam of molecules being attenuated not by annihilation but by particles being randomly scattered out of the initial beam. The probability that a created photon will survive random annihilation after travelling distance d*(f) along its mode is p(d*(f)). Its probability of survival at d*(f) obeys p(d*(f))= [exp(-d*(f)/(<d*(f)>$_Q$)] with <d*(f)>$_Q$ the photon mean-free-path, found by integration over all possible d*(f) reached after creation, weighted by exp(d*/(<d*(f)>$_Q$)], which results in <d*(f)>$_Q$. If classical wave mean power decay at frequency f predicted by Maxwell's equations is to match the mean decay rate of photon energies hf from creation the relation <d*(f)>$_Q$ =<d*(f)>$_{Cl}$ must be satisfied. This equality allows use of known optical wave indices n(f), k(f) to determine the mean-free-paths in a collection of photons and provides a statistical "bridge" between classical optics and quantum photonics.

The mean photon lifetime <τ*(f)> within an absorber becomes, since <d*(f)>$_Q$ =<d*(f)>$_{Cl}$

$$\langle \tau^*(f)\rangle = \frac{\langle d_a^*(f)\rangle}{c*(f)} = T\frac{\langle d_a^*(f)\rangle}{\lambda^*} = \frac{T}{4\pi k(f)} \qquad (2)$$

with $d_a^*(f)$ the variable distance travelled per photon per mode to its annihilation location. Period $T=1/f$ and $\lambda^*(f) = c^*(f)T$ the modal wavelength, and if $4\pi k(f) >1$ a photon is on average annihilated before completing one rotation cycle or travelling distance $\lambda^*(f)$. This occurs in select dielectric frequency bands, near local resonances and dominates when the bulk is plasmonic or $\varepsilon_1(f) <0$ as in phonon Restrahlen bands. Free photon modes exist in semiconductor band gaps but when thermally excited into these photons are not annihilated unless local defects are present. The density of occupied internal energy modes at $T_K$ are material dependent. The models we will now present are applied to spectral photon densities inside silver, germanium, water, and a cavity. They are quite different in magnitude and spectra. A closed cavity is the only "internal matter" where all photons avoid annihilation, which occurs in the wall where cavity photons are also created. Cavity modes are not continuum modes, but some can extend through a small hole in the wall far into the continuum.

Thermodynamics requires that photons cannot exist independently of their modes, so that planes of constant mean particle density and mode amplitude are always parallel to planes of constant modal phase. Solutions of classical wave equations however exist in which phase and amplitude evolve in different directions (5-7). Such classical solutions are not compatible with photonics. The solution that matches photon flows also defines exit refraction angles and was first established (8) at GHz frequencies. Those models can be used to find exit refraction angles for exit modes, hence their photons. Plots of $n(f)$ and $k(f)$ used in the following models are in the supplement at infra-red wavelengths occupied to varying degrees at temperatures in the range 300K to 500K. They are needed to derive and demonstrate the variations between internal photon spectral energy densities $\rho_U(f,T_K)= [dU(f,T_K)/df]$ for a dielectric (water), a semiconductor (germanium), a metal (silver), and the blackbody spectrum emerging from a closed cavity, where $n(f)=1, k(f)=0$. Experimental data (9, 10) which combine all spectral-directional emission profiles from smooth heated metal samples can be used to test these models, but as exactly modelled later the spectral intensities observed at normal exit for different $T_K$ provide the easiest validation route.

Our models for $\varepsilon_H$ and directional emittance spectra rely on internal interface reflectance not external reflectance. The non-photon excitations required for creation and annihilation can be moving phonons and excited band electrons, or localized defects on atoms, ions, molecules, and lattices. The internal density $P(f,T_K)$ of these excitations determines photon internal spectral density $N(f,T_K)$ in the steady thermal state. Together they set the available configurations $W(f,T_K)$ that define the photonic internal entropy flux density $S(f,T_K)$ at each frequency. A step-up in heating power $\Delta(dQ/dt)$ creates an initial transient that vanishes once a new steady state at a new $T_K$ exists, which includes a stable addition to photon emission. A fraction in sample volume V of $VN(f,T_K)$ photons are emitted and we limit maximum temperature so that $VP(f,T_K)$ are trapped. A time-averaged final equality between $N(f,T_K)$ and the portion of $P(f,T_K)$ active in photon creating and annihilating collisions results. Four power flows are then in balance at temperature $T_K$. They are (i) heat input rate $dQ/dt$ (ii) and (iii) summed rates of creation and annihilation of internal photon energy (iv) emitted radiant power $P_H(T_K)$. Detailed balance between input power and output power $P_H$ applies isothermally to all internally generated photons with a chance of escaping so this process is conservative with $\Delta(S(f,T_K)=0$, and thus is allowed by the 2nd Law, despite transforming heat input into some information on the host material. Lawrence(11) showed photoelectric and thermionic emission of electrons parallels the external detailed power balance between dQ/dt input, and $P_H(T_K)$ of thermally emitted photons. If annihilation at any frequency f is absent, detailed balance in the steady state is between that mode's take-up of a fraction of dQ/dt, creation of internal photon power at frequency f, and hemispherical power flow $P_H(T_K,f)$ in the external mode at f.

Each exit quantum mode depends on the internal rotational average of its incident photon fields projected onto the interface. A time-averaged normal reaction force results which depends on the time-averaged projection of particle fields **E**(f,t) and **H**(f,t) within each incident mode into the frame of reference defined by the local interface and its normal. In that frame photons in TE and TM modes experience different time-averaged forces which depend on the angle of internal incidence θ*. One reflected quantum mode and one transmitted quantum mode at each mean polarization results for each incident mode. The relative occupancy of the four modes generated by the interface, all at the incident energy, is governed by their different mode amplitudes relative to that of the incident mode. These determine the probability split for each incident TE and TM photon. These amplitudes can be found from the solutions of two time-independent Schrödinger equations, one for each time-averaged hence TE and TM reaction potential. Time-averaging over fluctuations in quantum transport properties thus allows alignment of classical power and mean photon power flows. Mishchenko (*12, 13*) recently argued such agreement was unlikely, but did not take account of the time-averaging that links classical and quantum power flows inside absorbers, nor the crossing of ground-state (empty) modes. Energy entropy leads to Shannon(*14*) information so correct entropy flows are important. The Planck(*2, 3*) cavity model will now be updated for photons generated within matter. The number of internal photon modes at frequencies up to f is given by equation (3) with two spin modes at each frequency. Each of these two modes is equally occupied internally so each contributes half the time-averaged internal flux at $T_K$.

$$N(f) = \frac{8\pi}{3} \frac{V}{[c^*(f)]^3} f^3 = \frac{8\pi}{3} \frac{V}{c^3} \frac{c^3}{[c^*(f)]^3} f^3 = \frac{8\pi}{3} \frac{V}{c^3} n(f)^3 f^3 \qquad (3)$$

The expressions for total intensities in each internal direction within matter follow from integration over occupied frequencies to yield values quite distinct from those in a cavity. From the sequence of equations (4) to (7) based on equation (3) the factor $(T_K)^4$ is retained. The integral of final interest is over the internal photon energy spectral densities $\rho(f,T_K)$ of equation (7). As for cavities the integral over f was transformed into one over the dimensionless variable x = (hf/kT) hence over $\rho(x(f))$. From equation (3) the number of modes dN(f) present between f to (f+df) within volume V becomes

$$dN(f) = \frac{8\pi V}{c^3} \left[ [n(f)]^3 f^2 + f^3 [n(f)]^2 \frac{dn(f)}{df} \right] df \qquad (4)$$

Occupation of these modes depends on the Bose-Einstein distribution function 1/[exp(hf/kT) -1] so the photon number spectral density becomes

$$\frac{1}{V} \frac{dN(f,T_K)}{df} = \frac{8\pi}{c^3} \frac{1}{\exp\left(\frac{hf}{kT_K}\right) - 1} \left\{ f^2 (n(f))^3 + f^3 (n(f))^2 \frac{dn(f)}{df} \right\} \qquad (5)$$

$dU(f,T_K)$ the photon contribution to internal energy density between frequencies f and (f+df) then becomes in dimensionless energy units $x=(hf/kT_K)$

$$\frac{dU(f,T_K)}{df} = \frac{8\pi k^3}{c^3 h^2} T_K^3 \left\{ \frac{x^3 (n(x))^3}{\exp(x) - 1} \left[ 1 + x \frac{d(\ln(n(x)))}{dx} \right] \right\} \tag{6}$$

Integration leads to the photonic contribution to internal energy with df= (kT$_K$/h)dx

$$U(T_K) = \frac{8\pi k^4}{c^3 h^3} T_K^4 \int_0^\infty dx \left\{ \frac{x^3 (n(x))^3}{\exp(x) - 1} \left[ 1 + x \frac{dln(n(x))}{dx} \right] \right\} = \gamma T_K^4 \beta_{CM} \tag{7}$$

$\beta_{CM}$ is the number resulting from the integral in equation (7) over internal spectral density $\rho(x(f))$. It replaces $\pi^4/15$ which occurs only when n(f) = n(x) =1. Each replacement number from equation (7) is specific to one material and sensitive to T$_K$. Figs. 1-3 show comparisons between the interior spectral densities in universal thermal intensity units of 8.7306x10$^{-9}$T$_K^4$ Wm$^{-2}$, expressed as a function both of x(f,T$_K$) and of external wavelength for water, germanium and silver at various T$_K$ up to 500K. For actual intensities these plots are thus multiplied at each λ by 8.7306x10$^{-9}$T$_K^4$. Internal spectral densities are seen to be qualitatively and quantitatively distinct for dielectrics, semiconductors and conductors. $\beta_{CM}$ thus defines the number of universal thermal units of intensity within each material, including inside a cavity where $\beta_{CM}$= $\pi^4/15$. In fig. 1(left) for water three strong photon resonances that multiply $\gamma T_K^4$ occur. That at 6.2 μm is due to the H-O-H bending mode, ~13 μm due to the libation, or whole H$_2$O vibrations, and ~21 μm due to H$_2$O molecules linked by hydrogen bonding(*15*). Each radiance intensity peak occurs at temperature independent wavelengths, but with temperature sensitive amplitudes. Many dielectrics display free photon local resonant features.

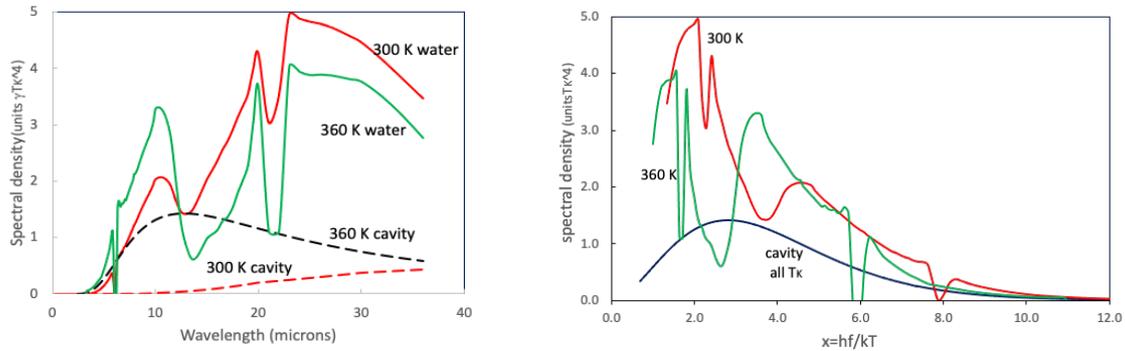

Fig. 1. Photon spectral density within water, and within a cavity at temperatures 360K (green) and 300K (red) as a function of wavelength and of x =hf/kT$_K$ using n(f), k(f) from Hale et al(*16*). The cavity spectral densities are fixed at all T$_K$ unlike those within matter.

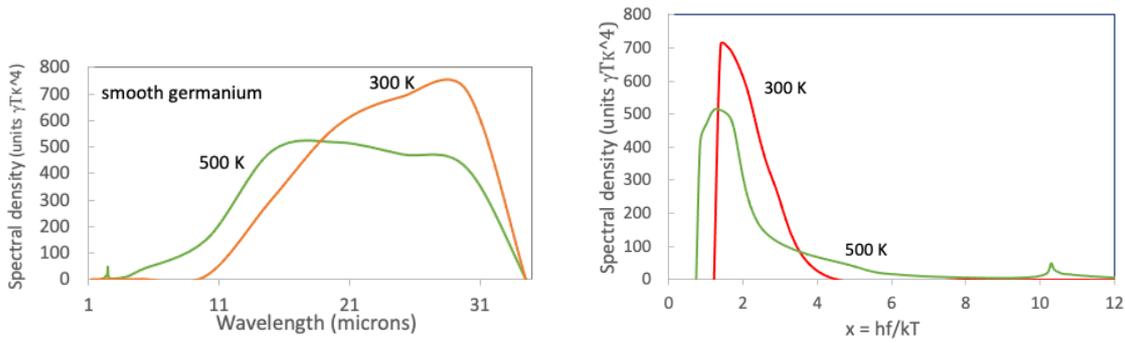

Fig. 2. Photon spectral density within germanium at temperatures 500K (green) and 300K (red) as a function of wavelength (left) and of x = hf/kT (right) using n(f), k(f) from Nunley et al(*17*).

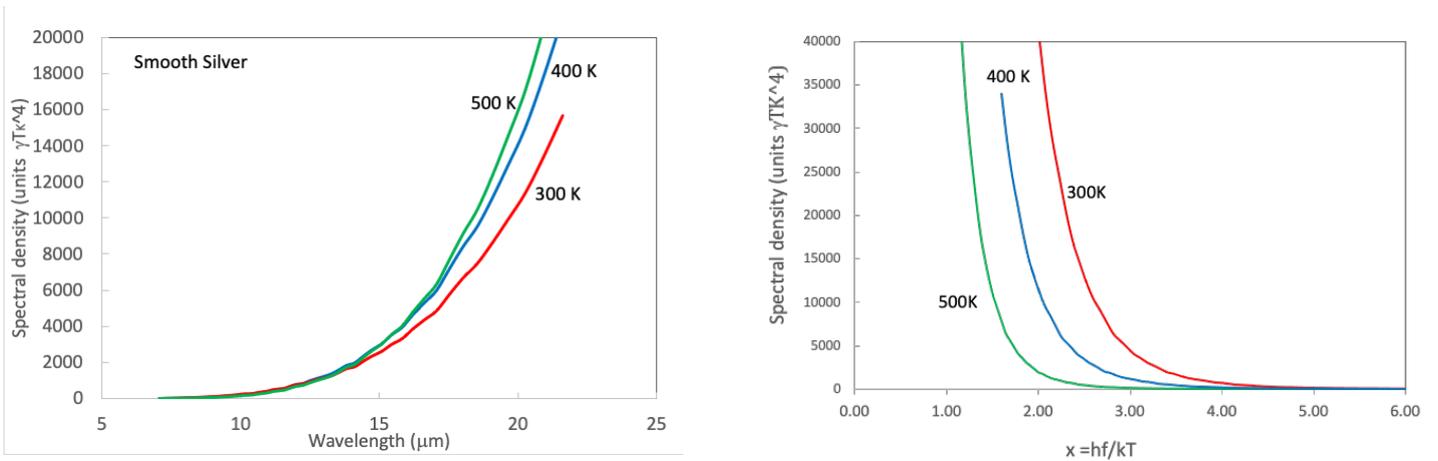

Fig. 3. Photon spectral density within silver at temperatures of 500K (green), 400 K(blue) and 300K (red) as a function of wavelength(left) and of x = hf/kT (right) with n(f), k(f) from Rakic et al(*18*) where maximum wavelength was 35 μm , hence the cut-offs (right) at short x.

The local resonances in Fig. 1 within a "free" photon system are analogous to the local resonances built-up out of free electron modes in noble metals doped with atoms whose local d-orbital energies overlap the host conductor's free electron s-band. Such hybrid electron states are known as Friedel(*19*) or Anderson(*20*) virtual-bound states. Internal photon modes form similar hybrid states with local absorption modes in dielectrics and fluid molecules. They are similar but different to "bound states in a continuum"(*21*). Shielded internal modes transport "free" photons of which some elastically enter linked external continuum modes and some of those have been delayed within hybrid resonant states. For semiconductors Fig. 2 shows a weak anti-resonance, or free mode exclusion near the band edge. An amorphous photonic lattice is indicated, which should vanish in a perfect crystal.

The internal spectral density plotted for water is not complete, along with that for pure silver. To complete these plots index data at wavelengths above 40 μm in water is needed, and well above the 25 μm limit in the silver data used(*18*). At long λ in conductors, frequency must drop below the electron relaxation rate for a complete plot and in pure silver this occurs around λ = 40 μm(*22*). Extended Drude models should also generate closed plots. The maximum spectral energy densities in $\gamma T_K^4$ units at 300 K inside silver are >2x10$^4$ compared to ~400 in germanium, ~4.6 inside still water, and just above 1 inside a cavity.

The sum of spectral energy densities yields each material's total internal photonic energy density $U(T_K)$ Jm$^{-3}$ and internal intensity $I(T_K)$ Wm$^{-2}$ passing through internal area A= $4\pi r^2$ with r the radius of any small internal sphere where select modes converge and cross as in fig. 4. Existing cavity-based energy density can be easily compared but more generally

$$U(T_K) = \beta_{CM}(T_K)[\gamma T_K^4]\, Jm^{-3} \tag{8}$$

Internal radiance elements passing through these small spheres are $d\Lambda_{int}(\theta^*,\phi^*,T_K)$= [$\gamma\beta_{CM}(T_K)T_K^4/4\pi M$] Wm$^{-2}$Sr$^{-1}$ so that $\gamma\beta_{CM}$ replaces $\sigma$ and M is a large integer such that M = $4\pi/d\Omega^m$ with $d\Omega^m$ the steradians enclosing common $d\Lambda_{int}$. The superscript m in $d\Omega^m$ represents each internal, rotated coordinate frame m =1 to M. $d\Omega^m$ is the fixed, smallest steradian element per frame being normal to each one's base plane and encloses fixed radiance $d\Lambda_{int}(T_K)$. From the areas under the plots in Figures (1-3) relative to those for the black body it is apparent that cavity $\beta_{CM} = \pi^4/15$ is less than all other $\beta_{CM}$ due to the increased mode density in matter. The spectral cavity density is amplified by factor $n(x)^3[1+xd\ln(x)/dx]$ from equation (6) at each x(f). The fraction of photons incident on an interface at each $(\theta^*,\phi^*)$ then emitted at each frequency is a function of internal reflectance $R(\theta^*,\phi^*,f)$ with $1-R(\theta^*,\phi^*,f) = \varepsilon(\theta^*,\phi^*,f)$ a directional emissivity. Reflectance applies differently to TE and TM modes which are equally occupied internally so $R(\theta^*,\phi^*,f)= 1/2[(R_{TE}(\theta^*,\phi^*,f) + R_{TM}(\theta^*,\phi^*,f)]$ and $(1-R(\theta^*,\phi^*,f))$ defines the total, now polarized, transmitted spectral flux. Only at $\theta^*=0°$ does $R_{TE}(\theta^*,\phi^*,f) = R_{TM}(\theta^*,\phi^*,f)$ and $R_{TE}(\theta^*,\phi^*,f)$ with $R_{TM}(\theta^*,\phi^*,f)$ are determined by n(f), k(f) and $(\theta^*,\phi^*)$.

Energy conservation in the split between interface reflected and transmitted fluxes at each frequency means 1- $R_{TE(TM)}(\theta^*,\phi^*,f) = T_{TE(TM)}(\theta,\phi;\theta^*,\phi^*; f)$ provides a power and intensity based link between $(\theta^*,\phi^*)$ and exit direction $(\theta,\phi)$ with $T_{TE(TM)}(\theta,\phi;\theta^*,\phi^*; f)$ each interface transmittance. Refraction intensity impacts are embedded in the interface transmittance value and show up in observed exit intensities of external TE and TM modes. When n(f) rises $\beta_{CM}$ increases and $1-R(\theta^*,\phi^*,f)$ falls. Radiative cooling rates thus depend on the increase in internal photon density and the fall in thermal emittance $\varepsilon(\theta^*,\phi^*,T_K)$ as detailed later, which results from summing $(1-R(\theta^*,\phi^*,f))$ weighted by the bulk spectral densities at each f and $T_K$. for each sample as detailed in eqns.(5-7).

The combination of refraction and an internal critical angle $\theta_C^*$ means external hemispherical radiance usually arises from $M_C < M$ internal common coordinate frames for each common radiance elements $d\Lambda^m(\theta^*,\phi^*,f,T_K)$ within each impacting hemisphere in Fig.4. Radiance elements $(M-M_C)d\Lambda^m(\theta^*,\phi^*,f,T_K)$ per incident hemisphere are projected into the solid angles defined within the interface frame. Those with $\theta_C^*<\theta^*<90°$ are totally internally reflected though some may enter evanescent surface modes. Each hemisphere, whose internal centers as in fig. 4, lie on or just below the interface emit hemispherical radiance $d\Lambda_H$. The emerging radiance in Wm$^{-2}$ Sr$^{-1}$ flows through multiple $d\Omega_H$ from each of many adjacent spheres arrayed over sample exit area with common normal. Total power lost is linear in exit area $\Delta A$ which becomes A if sample is smooth and flat. The formal derivation of $\varepsilon_{Q,H}$ the quantum total emittance and of $\Lambda_H(f,T_K)$ following shows that eqn. (8) applies to all matter including that from a cavity's small aperture where $\varepsilon_{Q,H}= \pi^4/15$.

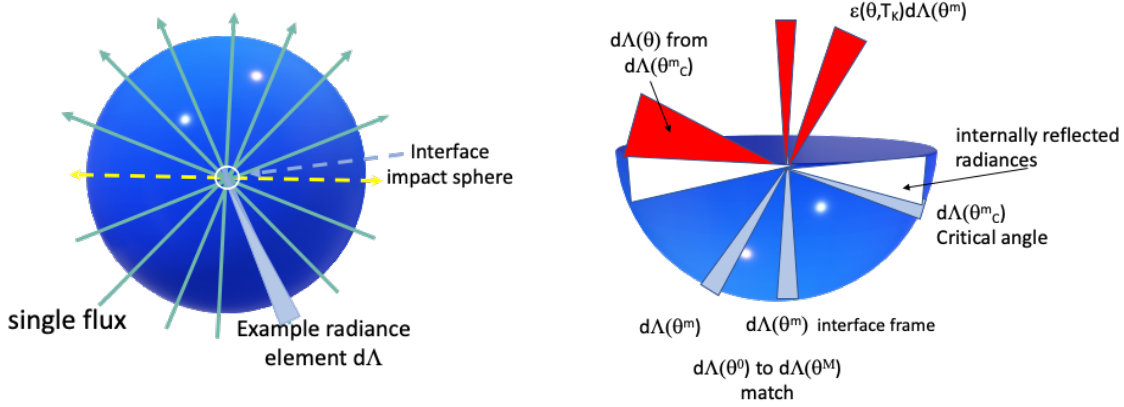

Fig. 4. Example sub-sets of internal and external modes, and select uniform internal radiance elements at finite $T_K$ (grey) and emitted $d\Lambda(\theta,\phi)$ (red). One ground state spherical set of modes passes through the small, circled internal sphere (left). Blue spheres are aids to symmetry visualization (they not physical). When the small sphere lies on or just below the interface each $d\Lambda^m$ from m=1 to $M_C$ are refracted differently. In the white zone $(M-M_C)d\Lambda^m$ are internally trapped.

$$\Lambda_H(T_K) = \frac{1}{2\pi}\varepsilon_{Q,H}(T_K)\gamma T_K^4 \quad Wm^{-2}\,Sr^{-1} \tag{9}$$

Refraction combines with material specific internal intensities after their reduction by interface-based reflectance to define exit intensity and exit radiance profiles at each frequency. Prediction and fitting of experimental data on emerging radiance and intensity profiles must thus account for the different intensity outcomes in each combination of exit direction, polarization and frequency. Modelled or measured exit profiles can be used to predict total loss rates and $\varepsilon_{Q,H}$. Due to refraction there are now two interrelated emissivities to consider, $\varepsilon(\theta^*,\phi^*,f)$ and $\varepsilon(\theta,\phi;\theta^*,\phi^*,f)$. The second emissivity defines what has traditionally been labelled simply $\varepsilon(\theta,\phi,f)$ since in traditional, surface source models (dictated by the Kirchhoff rule) only external flows were considered and $\varepsilon(\theta^*,\phi^*,f)$ did not exist. The dependence of each exit mode direction $(\theta,\phi)$ on internal mode direction $(\theta^*,\phi^*)$ must now be considered when modelling exit intensity directional profiles, but that step can be avoided if the prime interest is finding total output. Both $\varepsilon(\theta^*,\phi^*,f)$ and $\varepsilon(\theta,\phi;\theta^*,\phi^*,f)$ operate now on sample internal intensities.

If independent of $\phi$ or $\phi^*$ the relationship of interest becomes that between $\varepsilon(\theta^*,f)$ and $\varepsilon(\theta,\theta^*,f)$. It is fully developed in the supplement using Fresnel interface complex amplitude coefficients $r(\theta^*,f)$ for mode reflectance and $t(\theta,\theta^*,f)$ for each polarization mode. For example $\varepsilon_{TE}(\theta^*,f)=(1-|r_{TE}(\theta^*,f)|^2)$ and $\varepsilon_{TE}(\theta,f)=|t_{TE}(\theta^*,f)|^2$ since $t(\theta,\theta^*,f) = t(\theta^*,f)$ as it is expressed in terms of incident angle. The energy conservation relation for each TE mode resulting is $\varepsilon_{TE}(\theta^*,f)\cos\theta^*=\varepsilon_{TE}(\theta,\theta^*,f)\cos\theta$ and for TM $\varepsilon_{TM}(\theta^*,f)\cos\theta^*= \varepsilon_{TM}(\theta,\theta^*,f)\cos\theta$. Our emphasis in this paper is on internal photon spectral densities and resultant total power output, but measured directional spectral exit intensities can be used for model validation along with net cooling rates. Validation is simple at normal incidence where observed external intensities plus observed normal reflectance per reversed external mode lead directly to internal spectral densities per sample at each frequency. The complex Snell's Law(*7, 8*) can also be used in place of $r(\theta^*,f)$ and $t(\theta^*,f)$ to find θ after each step in θ*. Its basis in momentum conservation also ensures energy conservation for exit photons.

These relationships mean ρ(x(f,$T_K$)) or ρ(f,$T_K$) for internal spectral density in thermal intensity units γ$T_K^4$ as in eqns. (5-7) lead as detailed in the supplement to dI(θ,f,$T_K$)/(γ$T_K^4$)= ε(θ*,f)ρ(f,$T_K$)cosθ*= ε(θ,θ*,f)ρ(f,$T_K$)cosθ. For external radiance profiles dΛ(θ,f,$T_K$) relative to dΛ(θ*,f,$T_K$), refraction changes both exit cross-sections and solid angles. Accounting for both geometric changes leads to dΛ(θ,f,$T_K$)/(γ$T_K^4$) = ε(θ*,f)ρ(f,$T_K$)sinθ*cosθ*dθ* = ε(θ,f)ρ(f,$T_K$)sinθcosθdθ. The uniform internal power flow within d$I^m$[(θ*,ϕ*, f,$T_K$] and d$Λ^m$ [(θ*,ϕ*,f,$T_K$] for frames m =1 to M, when projected into the co-ordinate frame in which the interface is the base-plane fill its cross-sections and solid angles respectively. The sensitivity of exit refraction angles (θ,ϕ) to change in frequency at fixed internal incidence direction (θ*,ϕ*) were also considered when modelling spectral responses. Some exit directions were checked with the complex Snell's Laws for fluxes emerging from specific absorbers(*8, 9, 23*).

$ε_H(T_K)$ can now be based on internally incident radiance elements $dΛ_{int}$(θ*,ϕ*,$T_K$). M/2 common internal solid angles fill each basic incident hemisphere sketched in Fig. 4. Their common magnitude is d$Ω^m$=(d$ϕ^m$d$θ^m$) Sr. In the internal co-ordinate frame in which the interface is the base plane the number of d$Ω^m$ impacting within direction range θ* to (θ*+dθ*) and ϕ* to (ϕ*+dϕ*) are [sinθ*dϕ*dθ*] while dΩ(ϕ*,θ*)/2π is the fraction of each incident hemisphere in the interface frame occupied by dΩ(ϕ*,θ*). Thus [sinθ*dϕ*dθ*] common radiance elements (dΛ)*m strike the interface in each possible oblique incidence direction range. Schematics of a set of internal quantum modes passing through an example small bulk interior sphere (circled in white) were in Fig. 4. At an interface this small sphere's common crossing modes contain sequential photons which are randomly separated and gaps can be large.

Accuracy in radiative cooling rates, environmental radiance flows and risks to human thermal comfort are now very important. Replacing classical $ε_{Cl,H}$σ by $ε_{Q,H}$γ the quantum based factor, reveals a range of percentage errors when used to correct past calorimetric based $ε_{Cl,H}$ values. An expression for the ratio between the new and established total and directional emittances, intensities and radiance are derived in the supplement. Many careful past studies concluded $ε_{Cl,H}$ was either incorrect or their authors assumed unidentified error sources were present, for example (*24-27*). One of us some years ago carried out precise calorimetric studies(*10*) on spectrally selective solar absorbers. All had low emittance from their various thinly coated metal substrates. Our $ε_{Cl,H}$ from thermal data at that time based on σ$T_K^4$ gave results systematically above Planck-Kirchhoff predictions.

Planck's adoption of Lambert's external radiance model to define emissivity was thought to be justified by the Kirchhoff rule applied to reversal of all exit fluxes, whose projection added cosθ per exit direction. This forced Planck to add his extra factor of 2.0 to achieve Stefan's classical black-body intensities. A cavity aperture unfortunately for past models is the only case where there is no refraction and no basis for cosθ. Though internal radiance elements are still uniform but no longer universal, refracted and interface modified internal modes means that we can no longer use the Kirchhoff emissivity defined by reversed black body intensities as a basis for a cosθ intensity profile as detailed in reference (*28*), sections 2.4 and 2.5. From the example of another smooth metal silver in fig. 3 Lambert's spectral profile from his hot metal ribbon, correctly displayed a factor cosθ, but will have a very different spectral intensity distribution to that based on the Planck-Kirchhoff cavity-based approach. $\varepsilon_{Q,H}$ is derived next using $R_{TE}(\theta^*,\phi^*,f)$ and $R_{TM}(\theta^*,\phi^*,f)$ per mode acting on the common internal intensities. Surface topology varies local internal $R(\theta^*,\phi^*,f)$ across an interface so that emissivity and $\varepsilon_H$ will vary with surface location and its tilt where each mode crosses. Rough samples can be described statistically if the angular distribution of local normal is known. The smooth interface critical angle still applies for each combination of internal flux and local surface tilt. Surface gratings, thin patterned or uniform overlayers, dopants, nanostructures and multilayer thin film surface stacks can also provide specific local emissivity variations.

Various experimental proofs of the validity of the spectral densities in equations (4) to (7) are possible. The easiest uses FTIR data on spectral intensities emitted normally from warm or hot smooth samples. Internal photon spectral densities can be extracted from data as observed normal intensity is the product of observed normal spectral reflectance and internal spectral density with normal emission from smooth interfaces the only case where the Kirchhoff rule for emissivity remains correct. The Planck-Kirchhoff normal spectral response and that in our model have three factors in common the same external normal reflectance, the Bose-Einstein distribution function, and the factor $f^3$. The internal densities in eqns. (5-7) add an extra factor $n(f)^3[1+fdln(f)/df]$. The contribution of $n(f)^3$ to the normal spectral output amplifies the often weak spectral features from oblique external reflectance $R(\theta,\phi,f)$ using the Kirchhoff rule acting on cavity density. Model validation thus requires stronger spectral features in output from the factor $n(f)^3$ and a significant elevation of output intensities relative to cavity-Kirchhoff estimates, mitigated in part by internal reflectance., can show temperature sensitivity after scaling out the universal factor $\gamma T_K^4$ (see fig. 5 and the supplement) even when index $n(f)$ does not vary as $T_K$ changes

Both emissivities now act on $d\Lambda^m(\theta^*,\phi^*,f,T_K)$ to create an exit radiance element centered about exit angle θ after refraction. The interface area impacted by each set of hemispherical fluxes in fig. 4 has elemental area $r^2d\Omega^*$. Total output since the surface density of small spheres is $1/r^2d\Omega^*$ means $P_H(T_K)=[A\Delta P_H(T_K)]/(r^2d\Omega^*)$ with $\Delta P_H(T_K)$ the full output from one of the set of fluxes per hemisphere hitting the interface in Fig.4. Directional emittance $\varepsilon(\theta^*,\phi^*,T_K)$ follows from the weighted integration in equation (10) over all occupied internal modes for each internal direction. Emitted directional intensity and radiance usually fall for standard exit spreads due to refraction but negative refraction is possible among select incident photons that exit specific absorbers, especially when $k(f)$ is high. Important directional consequences result which do not occur classically. Each external radiance or intensity contains the power flows defined by impacting intensities resulting in occupied modes at $\theta(f)$ for each $\theta^*(f)$. Intensity lost per single internal TE mode is

$$I_{TE}(\theta,\phi,T_K) = \frac{8\pi k^4}{c^3h^3}T_K^4\,cos\theta^* \int_0^\infty dx(1-R_{TE}(\theta^*,\phi^*,x))\left\{\frac{x^3(n(x))^3}{\exp(x)-1}\left[1+x\frac{dln(n(x))}{dx}\right]\right\} \quad (10)$$

Directional emittances $\varepsilon_{TE}(\theta^*,\phi^*,T_K)$ is defined by the integration over frequencies and can be used to define emitted TE intensities from the internal integration over frequency or x in eqn. (10). To represent these emitted intensities in terms of exit direction θ, the result of the integration $\varepsilon_{TE}(\theta^*,\phi^*,T_K)$ can now be replaced with $\varepsilon_{TE}(\theta,\phi,T_K)[cos\theta/cos\theta^*]$ which leaves output TE intensity $\gamma T_K^4\varepsilon_{TE}(\theta,\phi,T_K)cos\theta=\gamma T_K^4|t_{TE}(\theta^*,f)|^2cos\theta$ as above and

proved in the supplement. Equivalent results follow for TM exit mode intensities. Summarising

$$I_{TE}(\theta,\phi,T_K) = \gamma T_K^4 \varepsilon_{TE}(\theta^*,\phi^*,T_K)cos\theta^* = \gamma T_K^4 |t_{TE}(\theta^*,\varphi^*,T_K)|^2 cos\theta \quad Wm^{-2} \quad (11)$$

TE radiance local directional output in Wm$^{-2}$Sr$^{-1}$ with axial symmetry reduces to

$$\Delta \Lambda_{TE}(\theta,\phi,T_K) = \gamma T_K^4 \varepsilon_{TE}(\theta^*,\phi^*,T_K)cos\theta^* sin\theta^* \Delta\theta^* = \gamma T_K^4 \varepsilon_{TE}(\theta,\phi,T_K)sin\theta cos\theta \Delta\theta \quad (12)$$

Using the number of common internal radiance elements $d\Lambda_{int}(\theta^m,\phi^m,T_K)$ projected into $d\Omega^* = sin\theta^* d\theta^* d\phi^*$ in the interface co-ordinate frame hemispherical emittance of eqn. (13) results with $\varepsilon(\theta^*,\phi^*,T_K) = \varepsilon_{TE}(\theta^*,\phi^*,T_K) + \varepsilon_{TM}(\theta^*,\phi^*,T_K)$ or $\varepsilon(\theta,\phi,T_K) = \varepsilon_{TE}(\theta,\phi,T_K) + \varepsilon_{TM}(\theta,\phi,T_K)$. Many production and processing techniques create interface topologies requiring biaxial or uniaxial interface reflectance or transmittance, so we retained explicit $\phi^*$ or $\phi$ dependence.

$$\varepsilon_H(T_K) = \frac{1}{2\pi}\int_0^{2\pi} d\Omega(\phi^*,\phi^*)cos\theta^* \varepsilon(\theta^*,\phi^*,T_K)$$

$$= \frac{1}{2\pi}\int_0^{2\pi} d\phi^* \int_0^{\pi/2} d\theta^* cos\theta^* sin\theta^* \varepsilon(\theta^*,\phi^*,T_K) \quad (13)$$

Water, snow and ice also reflect and scatter asymmetrically, after the impact of prevailing wind directions. The impact of each $\phi^*$ in emittance $\varepsilon(\theta^*,\phi^*,T_K)$ must often be retained. When internal reflectance R($\theta^*$, $\phi^*$,f) is independent of $\phi^*$ however equation (14) results.

$$\varepsilon_H(T_K) = \int_0^{\pi/2} d\theta^* sin\theta^* cos\theta^* \varepsilon(\theta^*,T_K) = \int_0^{\pi/2} d\theta sin\theta cos\theta \varepsilon(\theta,\theta^*,T_K) \quad (14)$$

With eqn. (12) eqns. (13) and (14) for $\varepsilon_H(T_K)$ show it acts on ($\gamma T_K^4$) not on cavity internal fluxes, nor on general internal intensities $\beta_{CM}(\gamma T_K^4)$. For R=0 however $\beta_{CM}=\pi^4/15=\varepsilon_{H,cav}$. Expressions for intensities and radiance exiting normally or spread over several $\theta$ directions can use eqns. (10-12). A repeatable, universal thermal radiation standard as an experimental reference for thermal emission is still needed. Cavity emission or the best available solid, "black-body" standards, as used in space missions(29), will continue to have a role. Internal density models and output however no longer depend directly on black-body intensities. Observable normal intensities as a function of $T_K$ using eqn. (10) are plotted for smooth silver and water in fig. 5. Normal internal and external reflectance match at each wavelength and are small at ~0.008 in smooth, pure Ag, while internal intensities exceed $45{,}000\gamma T_K^4$ Wm$^{-2}$. Normally emitted spectral intensities from water reproduce the strong resonance features of Fig. 1.

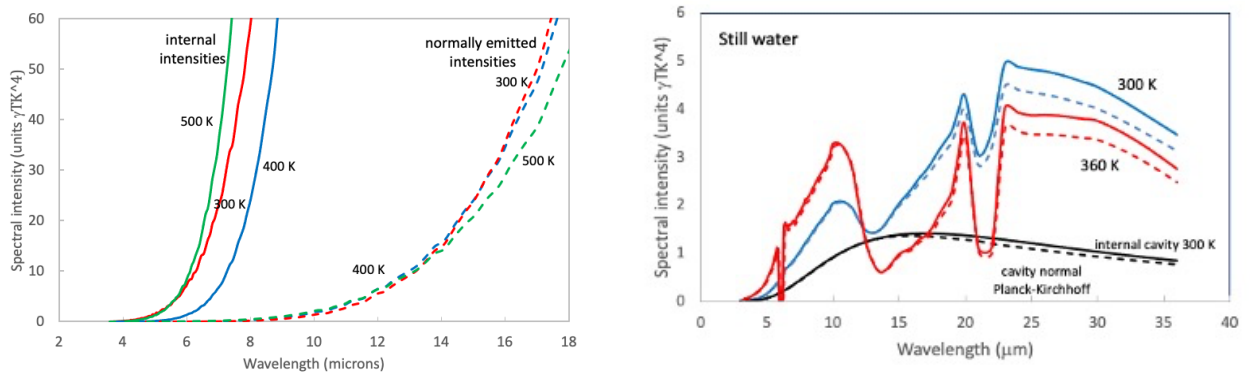

Fig. 5 Example normally emitted intensities (dashed) and internal densities, for smooth silver at 300 K, 400 K and 500 K (left) and still water at 300K and 360 K (right). Internal silver spectral intensities below $60\gamma T_K^4$ Wm$^{-2}$ are a small subset of those in fig. 3.

This project started when we realized that accurate solar cell cooling rates did not obey the values expected from the thermal emittance of their smooth glass covers(*28*). Our initial better fits had assumed that internal generation cancelled Lambert's external "cosθ" weighting for glass (*30*). Though correct for a cavity wall hole and approximately so for very diffuse (Lambertian) interfaces we later realized that smooth glass refraction was a feature in emitted flows. Refraction is the origin of Lambert's "cosθ" not his idea of surface based sources. Planck's photon density inside a cavity was ground-breaking but assuming Lambert's "cosθ" applied to emission from a "cavity" or black body led him to output intensities of $\sigma T^4/2$, half the classical value $\sigma T^4$. Had the cavity modes been treated as "pseudo-solid modes" in which photons were neither slowed nor annihilated, and output through a small hole is not refracted, $\sigma T^4$ would have emerged directly. Planck's rationale of 2-photon spins to justify his correction factor of 2.0 ignored the cavity density factor of $8\pi$ in equations (5-10) which already included both photon spins. Planck's extra 2.0 thus amounted to double counting of photon spin and the factor cosθ is due to refraction which drops out through a small cavity hole. Kirchhoff's $A(\theta,f) = \varepsilon(\theta,f)$ rule implied that emission at band-gap frequencies in semiconductors will not occur. In figure 2 and the supplement, it does occur as detailed balance requires its photon modes to be thermally occupied. There are often similarities for many samples between $\varepsilon_{Q,H}\gamma$ and $\varepsilon_{Cl,H}\sigma$ at limited $T_K$ provided Planck's extra 2.0 is included. An experimentally obvious weakness in the old models is their inability to predict the strong local hybrid resonances in Fig. 5 for water and that we have observed in FTIR spectra emerging from warm to hot dielectrics, for example from glass and silica near 10μm. These are due to the "virtual-bound state" hybrid resonances involving free photon internal modes overlapping some localized oscillator modes at the same energy.

Partial exit coherence exists in addition to the added information this paper's thermodynamic models allow input heat to generate without violating the 2nd Law. Our model does not predict the entangled coherence and incoherence within partial coherence so is still incomplete. The degree of partial coherence between the photons within a mode does not alter its energy content. A new source of entropy and quantum information, not defined by energy must therefore exist. Partitioning of excitations in terms of time thus needed consideration. Doing that implies classical, relativistic and Hawking's black-hole versions of time, as currently understood, need further refining based on quantum physics. Hawking realized this, and it became the focus of his later work (*31*) to explain the information paradox in a black-hole's thermal radiation. A related information paradox exists in the partial coherence within basic thermal radiance(*32*). Progress requires a statistical description of time and a physical mechanism that generates the required quantum time distribution. A corollary to the existence of dual quantum entropy is that the classical "arrow-of-time" and its inherent "information death" no longer exist. Instead, the cycles of random quantum creation, annihilation and the scope for new information to emerge when shifts in thermal or chemical environments occur, will go on indefinitely. Dual entropy also means the need for a Maxwell demon(*23, 33*) can be dispelled.

Acknowledgements
Discussions on aspects of the photonics, and thermodynamics required in this study occurred with Christopher Poulton and Michael Cortie. Recent studies by PhD researchers Marc Gali, Maryna Bilokur and Matthew Tai provided helpful thermal radiance data.

References
1. M. Planck, *Eight lectures on theoretical physics*. (Courier Corporation, 2012).
2. M. Planck, The Theory of Heat Radiation, English translation by M Morton, P. Blakiston's Son & Co, Philadelphia. (1914).
3. M. Planck. (Dover: New York, 1959).
4. F. Sears, G. Salinger. (Addison Wesley Pub. Co., Reading, Mass. USA).
5. J. Shen, H. Yu, J. Lu, Light propagation and reflection-refraction event in absorbing media. *Chin. Opt. Lett.* **8**, 111-114 (2010).
6. Q. Zhang, The refractive angle of light propagation at absorbing media interface. *Optik* **126**, 4387-4391 (2015).
7. S. Zhang, L. Liu, Y. Liu, Generalized laws of Snell, Fresnel and energy balance for a charged planar interface between lossy media. *Journal of Quantitative Spectroscopy and Radiative Transfer* **245**, 106903 (2020).
8. J. Garcia-Pomar, M. Nieto-Vesperinas, Transmission study of prisms and slabs of lossy negative index media. *Optics Express* **12**, 2081-2095 (2004).
9. W. Sabuga, R. Todtenhaupt, Effect of roughness on the emissivity of the precious metals silver, gold, palladium, platinum, rhodium, and iridium. *High Temperatures High Pressures(UK)* **33**, 261-269 (2001).
10. H. Willrath, G. Smith, A new transient temperature emissometer. *Solar Energy Materials* **4**, 31-46 (1980).
11. E. O. Lawrence, Transition probabilities: Their relation to thermionic emission and the photo-electric effect. *Physical Review* **27**, 555 (1926).
12. M. I. Mishchenko, Directional radiometry and radiative transfer: a new paradigm. *Journal of Quantitative Spectroscopy and Radiative Transfer* **112**, 2079-2094 (2011).
13. M. I. Mishchenko, Directional radiometry and radiative transfer: the convoluted path from centuries-old phenomenology to physical optics. *Journal of Quantitative Spectroscopy and Radiative Transfer* **146**, 4-33 (2014).
14. C. E. Shannon, A mathematical theory of communication. *Bell system technical journal* **27**, 379-423 (1948).
15. J.-B. Brubach, A. Mermet, A. Filabozzi, A. Gerschel, P. Roy, Signatures of the hydrogen bonding in the infrared bands of water. *The Journal of chemical physics* **122**, 184509 (2005).
16. G. M. Hale, M. R. Querry, Optical constants of water in the 200-nm to 200-μm wavelength region. *Applied optics* **12**, 555-563 (1973).
17. T. N. Nunley *et al.*, Optical constants of germanium and thermally grown germanium dioxide from 0.5 to 6.6 eV via a multisample ellipsometry investigation. *Journal of Vacuum Science & Technology B, Nanotechnology and Microelectronics: Materials, Processing, Measurement, and Phenomena* **34**, 061205 (2016).
18. A. D. Rakić, A. B. Djurišić, J. M. Elazar, M. L. Majewski, Optical properties of metallic films for vertical-cavity optoelectronic devices. *Applied optics* **37**, 5271-5283 (1998).
19. J. Friedel, ON SOME ELECTRICAL AND MAGNETIC PROPERTIES OF METALLIC SOLID SOLUTIONS. *Canadian Journal of Physics* **34**, 1190-1211 (1956).

20. P. W. Anderson, Local moments and localized states. *Reviews of Modern Physics* **50**, 191 (1978).
21. C. W. Hsu, B. Zhen, A. D. Stone, J. D. Joannopoulos, M. Soljačić, Bound states in the continuum. *Nature Reviews Materials* **1**, 1-13 (2016).
22. M. G. Blaber, M. D. Arnold, M. J. Ford, A review of the optical properties of alloys and intermetallics for plasmonics. *Journal of Physics: Condensed Matter* **22**, 143201 (2010).
23. P. Davies, What is life? *New Scientist* **241**, 28-31 (2019).
24. D. C. Hamilton, W. Morgan, "Radiant-interchange configuration factors NACA technical note 2836," (National advisory committee for aeronautics, 1952).
25. H. B. Curtis, Measurement of emittance and absorptance of selected materials between 280 deg and 600 deg K. *Journal of Spacecraft and Rockets* **3**, 383-387 (1966).
26. G. Abbott, N. Alvares, W. Parker, "Total Normal and Total Hemispherical Emittance of Polished Metals -Part 2," (NAVAL RADIOLOGICAL DEFENSE LAB SAN FRANCISCO CA, 1963).
27. T. Králík, V. Musilová, P. Hanzelka, J. Frolec, Method for measurement of emissivity and absorptivity of highly reflective surfaces from 20 K to room temperatures. *Metrologia* **53**, 743 (2016).
28. G. B. Smith, C.-G. S. Granqvist, *Green nanotechnology: solutions for sustainability and energy in the built environment*. (CRC Press, 2010).
29. M. Quijada, J. Hagopian, S. Getty, R. Kinzer, E. Wollack, *Hemispherical reflectance and emittance properties of carbon nanotubes coatings at infrared wavelengths*. SPIE Optical Engineering + Applications (SPIE, 2011), vol. 8150.
30. A. Gentle, G. Smith, Is enhanced radiative cooling of solar cell modules worth pursuing? *Solar Energy Materials and Solar Cells* **150**, 39-42 (2016).
31. S. W. Hawking, The information paradox for black holes. *arXiv preprint arXiv:1509.01147*, (2015).
32. S. Wijewardane, Y. Goswami, Exergy of partially coherent thermal radiation. *Energy* **42**, 497-502 (2012).
33. C. Elouard, D. Herrera-Martí, B. Huard, A. Auffeves, Extracting work from quantum measurement in Maxwell's demon engines. *Physical Review Letters* **118**, 260603 (2017).

# Supplementary Materials for

**The Stefan-Boltzmann constant revisited for photons thermally generated within matter**

G.B. Smith*, A.R. Gentle, M.D. Arnold, School of Mathematical and Physical Sciences, University of Technology Sydney; Broadway, NSW, Australia

*Corresponding author. Email: geoff.smith@uts.edu.au

This supplement covers experimental data used in the models in the main text, including relations and evidence that can be used in model validation, and brief discussion of why the Planck-Kirchhoff models gained long term experimental traction despite their thermodynamic and optical flaws that become apparent upon studying photon generation and annihilation internally. Main text models are extended here to include further details on how refraction effects influence exit intensity data. These models also show that Lambert's cosθ intensity profile was due to refraction not his original model of arrays of elemental surface emitters (*1*) (whose output is not refracted). The impact on intensity profiles of sample geometry, surface topology and the experimental optical set-up used to measure radiant intensity are briefly discussed.

# (A) Wave indices used to model internal photon spectral densities

The relations (S1), (S2) link host defined complex dielectric response functions $\varepsilon(f) = \varepsilon_1(f)+i\varepsilon_2(f)$ to internal wave indices $n(f)$, $k(f)$. In the main paper they are used to define both ground state quantum modes and Maxwell waves. Being photon based they define individual photon speeds in matter, and also depend within absorbers on time averages of lifetimes and distances reached by photons at annihilation from their creation location. These indices were used to show that the mean distance reached by classical wave power entering an absorber and the mean-free-paths of free internal photons match. As plotted in Figs S1, S2 and S3 they were then used to establish internal photon spectral densities. $n(f)^3$ is included in one of these plots and partially in another, as its dispersion is a primary influence on the internal spectral response and the spectral impact of temperature change.

$$n(f) = \frac{1}{\sqrt{2}}\sqrt{[(\varepsilon_1^2 + (\varepsilon_2)^2]^{\frac{1}{2}} + \varepsilon_1} \qquad (S1)$$

$$k(f) = \frac{1}{\sqrt{2}}\sqrt{[(\varepsilon_1^2 + (\varepsilon_2)^2]^{\frac{1}{2}} - \varepsilon_1} \qquad (S2)$$

Measured $n(f)$, $k(f)$ values for three common materials, water, germanium and silver, were chosen to represent typical infra-red internal photon transport in dielectrics, semiconductors, and metals respectively. These three matter classes have internal photon spectral densities which are qualitatively quite different. Only that in the dielectric example water are spectral trends on average parallel similar to those in radiance from a cavity, but these dielectric spectra are also overlaid with strong localized hybrid resonant features internally for otherwise free photons. Their distinct presence is strong in output but quite weak and barely seen in traditional models. The influential dispersion of $n(f)^3$ is plotted in full for water as it amplifies the local resonances in which "free" photon modes hybridise with local resonances. For germanium $n(f)^3$ maximises at 74.6 as maximum $n(f)$ is 4.21 and for silver at 12 μm $n(f)^3$ exceeds 4,000 and keeps rising at longer wavelength. Resonant features are obvious and well-known near 10 μm in emission from silica and glass, but are barely present in Planck-Kirchhoff cavity models. The emission spectra in the main text for water show three such strong resonances in $n(f)^3$ of Fig. S1. $n(f)^3$ raises all intensities to values l above those emitted using Kirchhoff emissivity operating on cavity output.

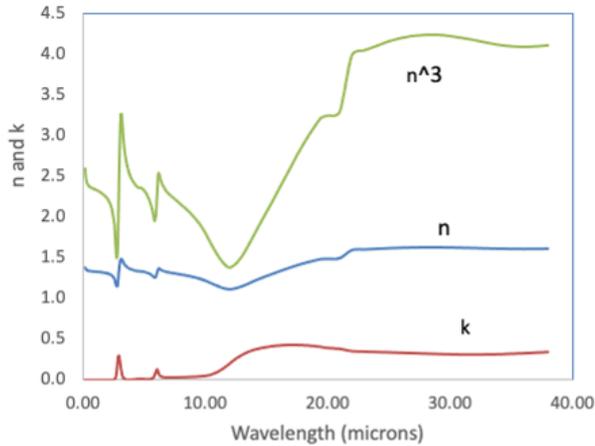

Figure S1. n(f) and k(f) inside water (based on n, k from Hale et al(2)).

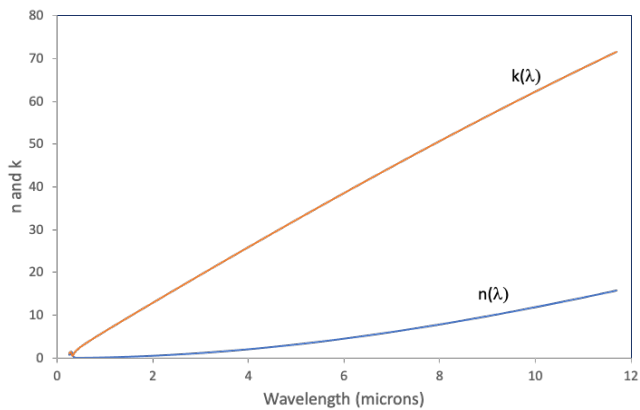

Figure S2. n(f) and k(f) inside silver (based on n(λ), k(λ) from Rakic($3$) ). n(λ)³ at 12 μm exceeds 4000. It plays a dominant role in Ag internal energy and exit intensities as seen in Figs. 3 and 5 main paper.

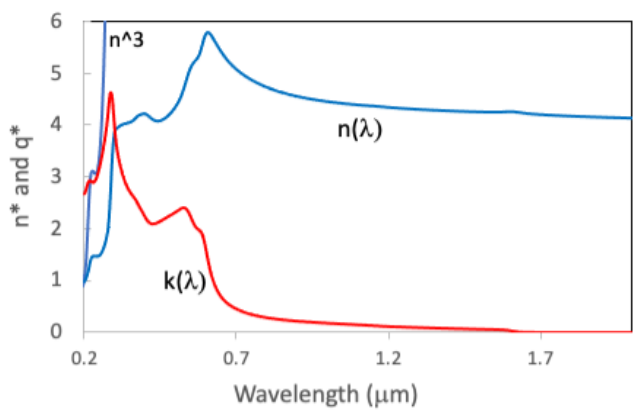

Figure S3. n(λ), k(λ) and part only of n(λ)³ inside germanium. n(λ), k(λ) data are from Nunley et al($4$). n(λ)³ exceeds 170 at its maximum value and levels off beyond 30 μm at about 67. k(λ) ~ 0 above 2.5 μm.

## (B) Normal spectral intensities from smooth, opaque examples

A widely accessible experimental approach to model validation was described in the main text. It relies on observation of normal infra-red spectral emission intensities combined with normal spectral reflectance from smooth, warm samples. Extension to oblique emission validation is also possible based on eqns. (10-12) main text with suitable FTIR instrumentation which we have set up for that purpose. Kirchhoff's use of external reflectance to define A(θ,λ) hence ε(θ,λ) can be applied only only for normal emission as exit refraction impacts obliquely incident photons. Example normal emission predictions for water and silver are in the main text, that for germanium follows here. Emitted spectral densities are expressed in number of universal units of radiant intensity γ($T_K$)$T_K^4$ Wm$^{-2}$ emitted at each wavelength using eqns. (11,12) main text, plus equation (S3) for normal reflectance $R_0$(f) off a smooth interface using indices n(f) and k(f). We assume that n and k do not change significantly in the range 300K to 360K in water and from 300 K to 500K in Ag and Ge. A change in $β_{CM}$($T_K$) hence internal energy density with $T_K$ does now occur while change in spectral density with $T_K$ from eqn. (10) main text means thermal emittance is different in its spectral and temperature sensitivities even in the absence of any $T_K$ dependent index changes. These will be seen if normal FTIR output is recorded at two or more well separated sample temperatures and then scaled by each γ($T_K$)$T_K^4$. These additional sensitivities to $T_K$ are important to the energy performance of some building products and current standards and will also impact urban heat island models which describe the thermal response of whole city precincts.

n(f) and k(f) used in these studies are in the previous section. Their role in normal reflectance, internally and externally, for smooth matter uses

$$R_0(f) = \frac{(1-n(f))^2 + k(f)^2}{(1+n(f))^2 + k(f)^2} \quad (S3)$$

Ideally a near black-body repeatable standard is available for radiance calibration at sample temperature. Intensity can still be referenced against Planck's cavity spectral density emerging through a small hole in the cavity wall at matching sample temperatures. Plots for classical cavity predictions were not included alongside the plots for predictions based on our internal photonic model for normal intensities in fig. S4 for germanium. Though finite they are too weak. Those for water and silver are in the main paper. For smooth germanium at 300 K and 500 K normal and internal spectral intensities are compared. For germanium the Planck-Kirchhoff prediction of normal emission intensities are so far below those arising from internal excitation models they are not plotted. At 300K in universal intensity units $\gamma T_K^4$ they range from 0.01 to 0.42 at the wavelengths in fig. S4. Model validation from semiconductors will thus be definitive.

For smooth samples where n(f) and k(f) have been previously established by accurate experimental optics, such as spectral ellipsometry or from spectral reflectance and absorptance data, the fitting of these two indices across all or parts of a normal or oblique emission spectral band at known $T_K$ provides a third approach to experimental validation and a new way of determining complex indices at emission wavelengths. Differences in polarization of emission for TE and TM intensities can also be used as outlined next.

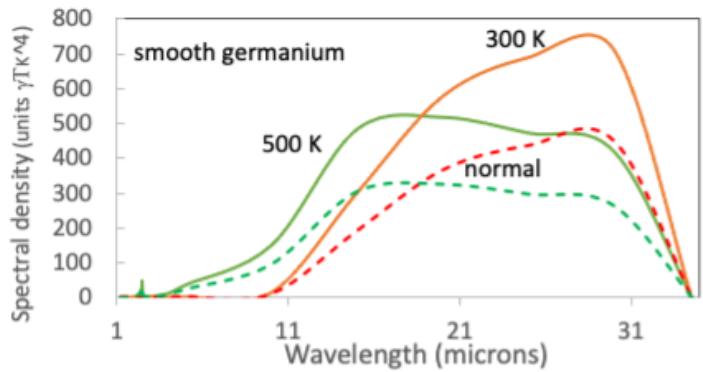

Fig. S4. Internal spectral density compared to normally emitted intensity (dashed) for smooth interfaces on germanium at 300 K and 500 K in $\gamma T_K^4$ units. For actual intensities multiply these plots at each λ by 8.7306x10$^{-9}$ $T_K^4$. The Planck-Kirchhoff prediction is relatively too close to zero for Ge to include. At 30 μm its normal output intensity prediction is $1.3\gamma T_K^4$ Wm$^{-2}$ against the $450\gamma T_K^4$ expected in this plot.

   (C) External intensity and radiance directional profiles compared to those from the Kirchhoff - Planck model

Interface refraction influences observed and modelled external radiance profiles for all exit directions except for normal emission. The experimental set-up used to monitor the intensity profile is also a consideration in intensity profile studies. Sample size, the view factor of the sample seen by the photon detection system, and background (non-sample) thermal radiance can each influence detected intensities and their profiles. Surface topology of each sample should also be accounted for in theoretical models and in observed intensity profiles. It is customary for structured interfaces to set a mean boundary, for example the smooth substrate below a rough or textured interface. A reference plane is needed to compare localized surface normal and is usually set by the sample mounting set-up. A distribution function for exit localized normal along such a reference plane for an interface sets each photon's exit direction. If an interface is textured with smooth nanoscale or microscale overlayer patterns, refraction angles and internal reflectance vary according to which material forms the local smooth interface.

For exit direction to the normal exceeding 15° to 25° for smooth or slightly rough liquid and solid matter the external refraction angle θ(f) grows rapidly as its θ*(f) internal mode approaches the critical angle $\theta_C(f)^*$. The additional conservation rule to that provided by the momentum and energy conservation from Snell's Law was outlined in the main paper. It led to a relation between internal property ε(θ*,f) and external ε(θ,θ*,f). We used the notation ε(θ,θ*,f) to indicate the link between θ and θ* but historically it has been labelled simply ε(θ,f) so that will be used from now on. The basic useful relation is ε(θ*,f)cosθ*= ε(θ,f)cosθ and ε(θ,f) is a function of incident angle θ* alone as we will now derive. Formulas for ε(θ*,f) and ε(θ,f) based on complex Fresnel mode amplitude coefficients r(θ*,f) and t(θ*,f) were presented in the main text. They are formally based on energy conservation at the interface involving **elastic tunnelling of photons at each polarization and frequency** into an allowed external mode. Optically this requires that TE and TM modes separately satisfy 1-R(θ*,f)= T(θ,θ*,f), with T(θ,θ*,f) each polarized transmittance into the continuum. The Kirchhoff rule however was based on the relationships A(θ,f)=1-R(θ,f)= ε(θ,f). But 1-R(θ,f)=T(θ*,θ,f) is not T(θ,θ*,f) as refraction of intensity transmitted out of opaque matter is T(θ,θ*,f)=|t(θ*,f)|²[cosθ*/cosθ] while T(θ,θ*,f) =|t(θ*,f)|²[cosθ/cosθ*] and the products |t(θ,θ*,f)|² and |t(θ*,θ,f)|² of each Fresnel

coefficient with its complex conjugate are identical. They are expressed in terms of n(f), k(f) and incident angle θ*. The Kirchhoff identity can thus only be used at normal external incidence. In addition we can now replace if desired (1-R(θ*,f)) from equation (10) main text with |t(θ*,f)|²[cosθ/cosθ*]. The expression for directional exit intensity of TE modes when independent of φ* thus becomes

$$I_{TE}(\theta,\phi,T_K) = \frac{8\pi k^4}{c^3 h^3} T_K^4 \cos\theta \int_0^\infty dx [t(\theta^*,x)]^2 \left\{ \frac{x^3 (n(x))^3}{\exp(x)-1} \left[ 1 + x \frac{d\ln(n(x))}{dx} \right] \right\} \quad (S4)$$

Equating this relation to that for $I_{TE}(\theta,\phi,T_K)$ of eqn.(10) main text and equating each modal flux making up the output integrals using the common internal density it is clear that energy conservation requires that TE modes satisfy

$$\frac{I_{TE}(\theta,f,T_K)}{\rho(f,T_K)\gamma T_K^4} = \varepsilon_{TE}(\theta^*,f)\cos\theta^* = \varepsilon_{TE}(\theta,f)\cos\theta \quad (S5)$$

with ε_TE(θ*,f) = 1- |r_TE(θ*,f)|² and ε_TE(θ,f) = |t_TE (θ*,f)|². TM emissivities can be established similarly. These equations provide an alternative to finding θ(f) to that from the complex Snell's Law as θ=cos⁻¹[(1-|r(θ*,f)|²)/|t(θ*,f)|²] for each internal incidence direction. From eqn. (S5) ρ(f,T_K)ε_TE(θ*,f) and ρ(f,T_K)ε_TE(θ,f) integrated over all f values leaves directional thermal emittances ε(θ*,T_K) and ε(θ,T_K) respectively and directional output intensities for TE modes are then

$$I_{TE}(\theta,T_K) = \gamma T_K^4 \varepsilon_{TE}(\theta^*,T_K)\cos\theta^* = \gamma T_K^4 \varepsilon_{TE}(\theta,T_K)\cos\theta \quad Wm^{-2} \quad (S6)$$

The combination of cavity modes modified by the emissivity from the Kirchhoff rule as the predictor of radiant output intensity has been used for over a century despite often being criticized, but the an alternative fundamental model has not been suggested until now. The need for models based on internal processes has been noted by a few authors(5, 6). Thermal response to heat input of all matter at finite temperature is also essential to the creation and destruction of information. The differences between predictions by the old model and its replacement, can be seen in the ratio in eqn. (S7) where internal photon generation within matter provides the numerator and $I_{TE,KP}(\theta,\phi,T_K)$ the intensity from the Kirchhoff-Planck approach is the denominator. This ratio is helpful to finding which parameters most influence differences in the two output predictions and under what conditions did the Kirchhoff-Planck approach approximate to the correct output intensities, despite its fundamental problems. Planck's addition of 2.0 to both total cavity and sample outputs did make rough agreement often possible after dividing calorimetrically determined cooling rates by $T_K^4$.

$$\frac{I_{TE}(\theta,f,T_K)}{I_{TE,KP}(\theta,f,T_K)} = \frac{\rho(f,T_K)\gamma T_K^4(1-R_{TE}(\theta^*,f))\cos\theta^*}{2.0\rho_{BB}(f,T_K)\sigma T_K^4(1-R_{TE}(\theta,f))\cos\theta} \quad (S7)$$

Cancelling common factors in the two spectral densities (which include the two Bose-Einstein thermal occupation factors) and using σ/γ = π⁴/15 leaves

$$\frac{I_{TE}(\theta,f,T_K)}{I_{TE,KP}(\theta,f,T_K)} = \frac{15}{\pi^4}\left[n(f)^3\left(1+f\frac{d\ln(n(f))}{df}\right)\right]\frac{[1-R_{TE}(\theta^*,f)]\cos\theta^*}{2.0[(1-R_{TE}(\theta,f))]\cos\theta} \quad (S8)$$

An equivalent result follows for TM modes. It is also of use to note that applying equation (S5) to both exit polarizations and $T_{TE}(\theta, \theta^*, f)$ and $T_{TM}(\theta, \theta^*, f)$ interface transmittances means

$$\frac{I_{TE}(\theta, f, T_K)}{I_{TM}(\theta, f, T_K)} = \frac{\varepsilon_{TE}(\theta, f)}{\varepsilon_{TM}(\theta, f)} = \frac{T_{TE}(\theta, \theta^*, f)}{T_{TM}(\theta, \theta^*, f)} = \frac{|t_{TE}(\theta^*, f)|^2}{|t_{TM}(\theta^*, f)|^2} \qquad (S9)$$

Emissivity labelled $\varepsilon_K(\theta,f)$ is from the Kirchhoff identity where absorptance $A(f)= 1-R(\theta,f) = \varepsilon_K(\theta,f)$ was defined for opaque matter using external reflectance, so $[1-R(\theta,f)]. \varepsilon_K(\theta,T_K)$. The Kirchhoff-Planck hemispherical power output per unit area then became

$$I_{H,K-P}(T_K) = 2.0\sigma T_K^4 \int_0^{\pi/2} d\theta \varepsilon_K(\theta, T_K) sin\theta cos\theta \qquad Wm^{-2} \qquad (S10)$$

From internal generation total exit intensity can now be expressed in two ways

$$I_H(T_K) = \gamma T_K^4 \int_0^{\pi/2} d\theta^* \varepsilon(\theta^*, T_K) sin\theta^* cos\theta^* = \gamma T_K^4 \int_0^{\pi/2} d\theta \, \varepsilon(\theta, T_K) sin\theta cos\theta \qquad (S11)$$

The use of Lambert's "$cos\theta$" factor for intensities and radiance weighting "$cos\theta sin\theta$" for all exit directions from smooth matter are thus correct, but due to refraction. A cavity or black-body radiator are the single exception as their outputs do not refract or reflect. For them $cos\theta = cos\theta^*$ and $\varepsilon(\theta,f)= \varepsilon(\theta^*,f)=1$ leaving uniform internal intensities $(\pi^4/15)\gamma T_K^4$ to escape without modulation. Lambert's surface source model for his hot ribbon profile data, was however adopted by Planck for a cavity. This forced the addition of factor 2.0 to all exit directions to achieve the classical black body exit intensity derived earlier by Stefan. Stefan's original classical black body intensity and our quantum intensity, like our use of complex indices from optics, provides another example where classical and quantum predictions merge. Had Planck realised that Lambert's "$cos\theta$" was due to interface refraction he would not have included "$cos\theta$" and not needed to add 2.0 to black body intensities to achieve the classical result. The cavity aperture is the sole exception to the presence of a $cos\theta$ factor in output intensity.

Finally we note that surface waves involving surface phonons, plasmons or other surface excitations, as often found in the presence of nano and micro-structures(7), can modify our outputs by creating thermal and non-thermal radiant output while travelling across an interface. Worth noting also is that thermally generated radiant fluxes that exit then impact a remote detector tilted at an angle $\theta_D$ to an interface normal, are a function of the area of the emitter $A(\theta_D)$ viewed. Practical signals may involve lens or mirrors, for the dependence of $A(\theta_D)$ on $\theta_D$. Each change if any in $A(\theta_D)$ as $\theta_D$ rises means the number of photons falling onto the detector as a function of $\theta_D$ changes. If $A(\theta_D)$ extends beyond the hot sample, so that two areas with quite different temperatures $T_1 \gg T_2$, are being viewed, with $T_2$ a much colder background detected photons can be dominated by sample area as $\theta_D$ rises. The area $A(\theta_D,T_1)$ may not change if any view expansion with tilt arises from growth in $A(\theta_D,T_2)$. Then the number of photons detected is dominated by $N(\theta_D)$ and is effectively fixed if the ratio $T_1^4/T_2^4$ is large. As $\theta_D$ changes the flux hitting the detector $N(\theta_D)$ is then approximately constant. Its projection normal to the detector becomes $N(\theta_D)cos\theta_D$ which equals $N(0)$ for normally emitted photon fluxes. View factor however was not the origin of Lambert's $cos\theta$ profile, but may be present in some outdoor radiance data where oblique views of multiple materials at a range of distances and viewing angles contribute to the signal. Projected area variations can also lead to more complex intensity profiles as tilt changes.

(C) Radiance from internal directions not accessible to classical waves introduced from external sources

Also relevant to surface waves we should note that thermal radiation based on internal generation can occupy internal modes beyond the critical angle. This cannot occur for input via smooth interfaces from external sources. Some modes in this range will create evanescent modes along the interface which have interesting thermodynamic and optical consequences. Another feature inherent to select absorbing matter is that a range of exit refraction angles were found which are less than the internally incident angle. If $\theta(f) < \theta^*(f)$ negative refraction at some frequencies is present, while negative n(f) is not involved. Examples are provided in references (*8*) and (*9*). Our models for silver and some water frequencies also yield such refraction outcomes in select bands and they can be expected in other matter. Such refraction outcomes in water and other matter require specific $\theta^*$ and large k(f) above or below n(f). Negative refraction outcomes when they occur are thus sensitive to two parameters $\theta^*$ and the relative magnitudes of k(f) and n(f). Observed thermal radiation profiles can then accumulate in select exit direction ranges, another feature that appears to be unique to the physics of internal quantum modes. Very hot matter up to sun and star temperatures thus needs further study in this regard, noting that the van Cittert-Zernicke idea(*10*) for narrow spread and ordered flows at remote distances from a complex source is based on classical wave interference.

Supplement references

1. J. H. Lambert, *Photometria, sive de Mensura et Gradibus Luminis, Colorum et Umbrae*. (Augsburg, 1760).
2. G. M. Hale, M. R. Querry, Optical constants of water in the 200-nm to 200-μm wavelength region. *Applied optics* **12**, 555-563 (1973).
3. A. D. Rakić, A. B. Djurišić, J. M. Elazar, M. L. Majewski, Optical properties of metallic films for vertical-cavity optoelectronic devices. *Applied optics* **37**, 5271-5283 (1998).
4. T. N. Nunley *et al.*, Optical constants of germanium and thermally grown germanium dioxide from 0.5 to 6.6 eV via a multisample ellipsometry investigation. *Journal of Vacuum Science & Technology B, Nanotechnology and Microelectronics: Materials, Processing, Measurement, and Phenomena* **34**, 061205 (2016).
5. J. Agassi, The kirchhoff-planck radiation law. *Science* **156**, 30-37 (1967).
6. C. Lessig, E. Fiume, M. Desbrun, On the Mathematical Formulation of Radiance. *arXiv preprint arXiv:1205.4447*, (2012).
7. H. Högström, S. Valizadeh, C. G. Ribbing, Optical excitation of surface phonon polaritons in silicon carbide by a hole array fabricated by a focused ion beam. *Optical materials* **30**, 328-333 (2007).
8. J. Garcia-Pomar, M. Nieto-Vesperinas, Transmission study of prisms and slabs of lossy negative index media. *Optics Express* **12**, 2081-2095 (2004).
9. S. Zhang, L. Liu, Y. Liu, Generalized laws of Snell, Fresnel and energy balance for a charged planar interface between lossy media. *Journal of Quantitative Spectroscopy and Radiative Transfer* **245**, 106903 (2020).
10. P. H. van Cittert, Die wahrscheinliche Schwingungsverteilung in einer von einer Lichtquelle direkt oder mittels einer Linse beleuchteten Ebene. *Physica* **1**, 201-210 (1934).